\documentclass[10pt]{article}

\usepackage{url} 

\usepackage{amssymb}

\def \ccomma{\raise 2pt\hbox{,}} 
\def \D {\hbox{d}}
\def \PII    {{\rm P_{\rm II}}}
\def \PI     {{\rm P_{\rm I}}}
\def \Ai {\mathop{\rm Ai}\nolimits}

\begin{document}

\title{On the Gurevich-Pitaevskii solution of KdV}

\author{Robert Conte
{}\\
\\ 1. Universit\'e Paris-Saclay, ENS Paris-Saclay, CNRS,
\\    LRC MESO, Centre Borelli, F-91190 Gif-sur-Yvette, France
\\
\\ 2. Department of mathematics, The University of Hong Kong,
\\ Pokfulam, Hong Kong.
\\
\\    E-mail Robert.Conte@cea.fr, ORCID https://orcid.org/0000-0002-1840-5095
{}\\
}

\maketitle

\hfill 

\begin{abstract}
The universal solution of the Korteweg-de Vries equation (KdV)
introduced by Gurevich and Pitaevskii
in order to describe the onset of dispersive shock waves
is known to also obey 
the self-similar reduction of the next member in the KdV hierarchy.
We show that, if this common solution obeys some lower order partial differential equation,
its differential order must be one,
and we provide its local representation as a converging Laurent series depending on both space and time.
\end{abstract}

\bigskip


\medskip

\medskip


\baselineskip=12truept 

\tableofcontents

\section{Introduction}

As shown by Gurevich and Pitaevskii (GP) \cite{Gurevich-Pitaevskii},
near the onset of their breaking,
dispersive shock waves are generically
described
by a particular, universal solution of the Korteweg-de Vries (KdV) partial differential equation (PDE)
\begin{eqnarray}
& &
u_t + u u_x + u_{xxx}=0.
\label{eqKdV3} 
\end{eqnarray}
The breaking is described by the cusp equation \cite[Eq (34)]{Gurevich-Pitaevskii}
\begin{eqnarray}
& & u^3 - t u + x=0,
\label{eqcusp} 
\end{eqnarray}
which has indeed only one real root $u$ before the breaking time $t=0$, 
while it has three real roots for $t>0$,
see \cite[Fig.~6]{Gurevich-Pitaevskii}.
Although this cusp equation is only derived in the dispersionless limit,
just after the breaking, dispersion regularizes this gradient catastrophe
and oscillations start filling the region,
see details in \cite[\S 3]{Gurevich-Pitaevskii}.

As confirmed numerically \cite{BMP1990} 
and theoretically \cite{Moore1990},
this solution is the unique 
smooth (i.e.~without poles on the real axis) real solution of KdV obeying 
the boundary conditions
\begin{eqnarray}
& & u \to -x^{1/3}, t \to \pm \infty.
\label{eqBoundaryCondition}
\end{eqnarray}
This Gurevich-Pitaevskii special solution has later been shown by Suleimanov \cite{Suleimanov-1994-ode4}
to also obey a fourth order nonlinear ordinary differential equation (ODE)
\cite{Suleimanov-1994-ode4} 
\begin{eqnarray}
& &
u_{xxxx} + \frac{5}{6} (2 u u_{xx} + {u_x}^2) + \frac{5}{18} (u^3 - t u + x)=0,
\label{eqF-V-PDE} 
\end{eqnarray}
in which the last three terms are precisely those of the cusp equation.

Indeed, considering this latter ODE as a PDE, both PDEs (\ref{eqKdV3}) and (\ref{eqF-V-PDE})
are compatible, in the sense that
\begin{eqnarray}
& &
(u_t \hbox{ taken from (\ref{eqKdV3})})_{xxxx} =
(u_{xxxx} \hbox{ taken from (\ref{eqF-V-PDE})})_{t}
\label{eqCompatibility}
\end{eqnarray}
is identically zero \textit{modulo} the two PDEs.

A summary of the properties of this quite important solution
can be found in 
Refs.~\cite{GST-2010,Opanasenko-Ferapontov-Abel,Suleimanov-Shavlukov-2021},
and in the recent review \cite{Kamchatnov-2021-review}.

\textit{Remark}.
Equation (\ref{eqF-V-PDE}) is identical, up to rescaling, 
to the particular case $\beta=0$ of an ODE first considered by Kitaev \cite{KitaevSainteAdele}
and denoted F-V in the classification of Cosgrove \cite{CosPole2}
\begin{eqnarray}
& & \hbox{F-V}:\
-v''''+20 v v'' +10 {v'}^2 - 40 v^3 + \alpha v + \kappa x + \beta=0.
\label{eqF-V-ODE}
\end{eqnarray}
This F-V ODE is itself a self-similar reduction of KdV5 (the next member in the hierarchy of KdV),
it belongs to the hierarchy of the first Painlev\'e equation \cite{KitaevSainteAdele}
and it is one of the very few fourth order ODEs \cite[\S A.3.6]{CMBook2} which possibly define 
a transcendental function irreducible to one of the six Painlev\'e functions.

Since the GP solution obeys two differential equations (one partial, one ordinary),
it is natural to investigate whether it obeys a ``smaller'' differential equation,
i.e.~one which would admit the two previous ones as differential consequences.
The problem addressed here is therefore to 
characterize this GP solution by the boundary condition (\ref{eqBoundaryCondition}) and
a ``lower order'' equation, if one exists,
i.e.~either by some PDE at most of second order, or by some ODE at most of third order.
After recalling previous results,
we first give two negative results about this common solution,
then a positive answer to the question, which is however only local.

\vfill\eject
\section{A perturbative result}
\label{section-Kudashev}

In Ref.~\cite{GST-2010},
the authors require the asymptotic series, 
\begin{eqnarray}
& & u(x,t)=t^{1/2} \sum_{j=0}^{+\infty} U_j(\Phi,z) 
 t^{-7 j/4},
z=x t^{-3/2}, 
\Phi=t^{-7/4} f(z)+s(z),
t \to \infty,
\label{eq-Asymptotic}
\end{eqnarray}
to obey both equations (\ref{eqKdV3}) and (\ref{eqF-V-PDE}).
This generates, at each expansion order,
a set of two PDEs, which must be compatible, for the coefficients $U_j$.
We will not repeat here this classical computation, 
but only state its main results, namely:
\begin{description}
\item
-- an elliptic dependence of $U_0$ on $\Phi$,
with $z$-dependent coefficients,
\item
-- a constant value $s(z)=\pi$ for the phase shift $s(z)$,
\item
-- and, most importantly,
a universal (i.e.~not depending on any parameter)
first order first degree nonautonomous ODE \cite[Eq (2.5)]{GST-2010} for $f'(z)/f(z)$,
\begin{eqnarray}
& & \frac{\D R}{\D z}=\frac{486 R^4-171 R^2+9 z R+5}{9(54 R^3 - 9 R+z)(2 R+3z)},
R=\frac{7 f}{4 f'}-\frac{3}{2}z.
\label{eq-Kudashev}
\end{eqnarray}
\end{description}
First obtained by Kudashev according to Ref.~\cite{GST-2010},
this ODE is invariant under $(R,z) \to (-R,-z)$
and, as proven in \cite{Opanasenko-Ferapontov-Abel},
its general solution admits a parametric representation 
\cite[Eqs.~(8)--(9)]{Opanasenko-Ferapontov-Abel}
in which $R$ and $z$ are algebraic transforms of a hypergeometric equation.

This remarkable result is only perturbative,
and, roughly speaking,
what we are looking for in this paper is a nonperturbative version of this result.

At present time, it is not known whether the F-V ODE (\ref{eqF-V-ODE}) is or is not reducible 
(in the sense of the theory of the explicit integration of ODEs
\cite{Umemura-Angers}),
therefore both cases must be examined.

In the next two sections, we first rule out the possibility for this unknown lower order equation to be an ODE.

\section{A first negative result}
\label{section-FV-irreducible}

If the fourth order ODE F-V is irreducible and if the common solution to (\ref{eqKdV3}) and (\ref{eqF-V-PDE}) obeys some ODE
of a strictly lower order than four
whose coefficients depend on $t$,
this contradicts the irreducibility of F-V,
therefore one concludes to the nonexistence of such an ODE of order at most equal to three.

\section{A second negative result}
\label{section-FV-reducible}

Assume now the F-V ODE (\ref{eqF-V-ODE}) to be reducible,
and assume the common solution to (\ref{eqKdV3}) and (\ref{eqF-V-PDE}) to obey some ODE
of order at most three
whose coefficients depend on $t$.

Then this ODE is necessarily either one of the reductions of KdV to an ODE,
or another ODE admitting (\ref{eqF-V-ODE}) as a differential consequence.
In the first hypothesis,
there exist three such reductions $(u,x,t) \to (U,\xi)$,
respectively to 
an elliptic ODE,
the first Painlev\'e equation $\PI$,
the 34-th equation of Gambier \cite{GambierThese} \cite{FA1981} (birationally equivalent to the second Painlev\'e equation $\PII$)
\begin{eqnarray}
& &
\left\lbrace
\begin{array}{ll}
\displaystyle{
u(x,t)=c-\frac{1}{3}U(\xi), \xi=x-c t, (U'' - 6 U^2)'=0,           
}\\ \displaystyle{
u(x,t)=-\frac{1}{3} (U(\xi)-t), \xi=x-6 t^2, (U'' - 6 U^2 - \xi)'=0,
}\\ \displaystyle{
u(x,t)=-\frac{1}{6} (3 t)^{-2/3} (U(\xi) - \xi/2), \xi= x (3 t)^{-1/3}, (2 U U'' - {U'}^2 - 4 U^3 + 2 \xi U^2)'=0,
}
\end{array}
\right.
\label{eq-KdV-reductions} 
\end{eqnarray}
but none of them contains a parameter which could be identified to $t$,
which rules out the first hypothesis.

In the second hypothesis, 
excluding now a reduction of KdV,
the reducibility assumption implies the existence of a first integral of 
(\ref{eqF-V-ODE}).
This first integral should by definition be independent of $\partial_t$,
but we have already seen that the elimination of $\partial_t$,
done in (\ref{eqCompatibility}), does not generate any new information,
so this second hypothesis is also ruled out.


{}From these two negative results, one concludes that, 
if the common solution to (\ref{eqKdV3}) and (\ref{eqF-V-PDE})
obeys some differential equation,
it is necessarily a PDE.
Let us now determine its differential order.

\section{A local answer}
\label{section-Laurent}

The KDV PDE admits a single, converging, Laurent series with two arbitrary coefficients \cite{WTC}.
In the notation of the invariant Painlev\'e analysis 
\cite{Conte1989} 
\cite[\S 4.4.1]{CMBook2},
this series is
\begin{eqnarray}
& &
\begin{array}{ll}
\displaystyle{
u= \chi^{-2} \left[-12+ (C-4 S) \chi^2 + (S-C)_x \chi^3 + u_4 \chi^4
\right.}\\ \displaystyle{\left.
+\left(\frac{1}{6} (C_t+C C_x + S C_x + C_{xxx} + S S_x - S_{xxx}) - u_{4,x} \right) \chi^5 
+u_6 \chi^6 + 0(\chi^7) \right],
}
\end{array}
\label{eqLaurentKdV} 
\end{eqnarray} 
in which $u_4$ and $u_6$ are two arbitrary functions of $(x,t)$.
In the above expression, the expansion variable $\chi(x,t)$, equivalent to the function $\varphi$
which defines the singular manifold equation $\varphi(x,t)=0$,
is only defined by its gradient,
\begin{eqnarray} & &
\chi_x =1 + \frac{S}{2} \chi^2,\
\chi_t= - C + C_x \chi  - \frac{1}{2} (C S + C_{xx}) \chi^2,
\label{eqChixt}
\end{eqnarray}
in which $(S,C)$ are two functions of $\varphi$ invariant under homographies,
\begin{eqnarray}
& &
S=\lbrace \varphi;x \rbrace
=\frac{\varphi_{xxx}}{\varphi_x}
 - \frac{3}{2} \left(\frac{\varphi_{xx}} {\varphi_x} \right)^2,
\label{eqS}
\\
& &
C=- \varphi_t / \varphi_x,
\label{eqC}
\end{eqnarray}
linked by the cross-derivative condition $(\varphi_t)_{xxx}=(\varphi_{xxx})_t$
identically satisfied in terms of $\varphi$,
\begin{eqnarray}& &
 S_t + C_{xxx} + 2 C_x S + C S_x = 0.
\label{eqCrossXT}
\end{eqnarray}

The requirement that the above series also obeys (\ref{eqF-V-PDE}) determines the values of the two arbitrary coefficients,
\begin{eqnarray}
& &
\left\lbrace
\begin{array}{ll}
\displaystyle{
u_4=- \frac{t}{12} + \frac{C^2}{4} + \frac{S^2}{5} + \frac{C_{xx}}{2} - \frac{S_{xx}}{5}\ccomma
}\\ \displaystyle{
}\\ \displaystyle{
u_6=\frac{5}{1008}x + \frac{5}{252} t C - \frac{5}{72} C^3 + \frac{t}{36} S
}\\ \displaystyle{\phantom{1234}
}\\ \displaystyle{\phantom{1234}
+ \frac{-2 C^2 S + 2 C_x^2 -4 C_{xt} + 2 C C_{xx} -4 S C_{xx} + C_{xxxx} - C_x S_x}{24}
}\\ \displaystyle{\phantom{1234}
- \frac{16}{315} S^3 + \frac{71}{1680} S_x^2 + \frac{8}{105} S S_{xx} 
- \frac{S_{xxxx}}{210}\cdot
}
\end{array}
\right.
\label{eqLaurent-SC} 
\end{eqnarray} 
The higher coefficients $u_j, j>6$ of the series (\ref{eqLaurentKdV})
now only depend on $(S,C)$
and one might wonder whether, 
when inserted in (\ref{eqF-V-PDE}),  
they put any additional constraint on $(S,C)$, i.e.~on $\varphi$.
This is not the case, because of the commutativity property (\ref{eqCompatibility}).
To confirm this, we have independently checked it up to $j=9$ included.
Therefore, if the common solution to (\ref{eqKdV3}) and (\ref{eqF-V-PDE})
obeys some differential equation,
it is necessarily a first order PDE.

Let us give a concrete example of such a possibility.
The nonlinear first order nonautonomous PDE
\begin{eqnarray} & &
v_x v_t - 4 x v^3=0, 
\label{eq-Example}
\end{eqnarray}
admits the explicit particular solution
\begin{eqnarray} & &
 v=\frac{1}{2 t (x^2+1)},   
\label{eq-Example-sol}
\end{eqnarray}
and a Laurent expansion 
without any arbitrary parameter, like ((\ref{eqLaurentKdV}), (\ref{eqLaurent-SC})),
\begin{eqnarray}
& & {\hskip -10.0truemm}
\begin{array}{ll}
\displaystyle{
v= -\frac{C}{x} \chi^{-2} 
-\frac{x C_t - 3 x C C_x + C^2}{4 x^2 C}\chi^{-1} 
}\\ \displaystyle{\phantom{12}
+ \frac{1}{192 x^3 C^3} \left[
x^2 (7 C^2 C_x^2 - 26 C C_t C_x +15 C_t^2 -56 C^3 C_{xx} - 64 C^4 S +32 C^2 C_{xt} - 8 C C_{tt})
\right.}\\ \displaystyle{\left.\phantom{1234567890}
+x (30 C^3 C_x - 18 C^2 C_t )
- 9 C^4
\right] \chi^0 
+ 0(\chi^1).
}
\end{array}
\label{eq-Example-Laurent} 
\end{eqnarray} 
This example only illustrates the possibility for a first order PDE
to admit at the same time an explicit solution and a zero-parameter Laurent series.

A natural question is now to match the asymptotic expansion (\ref{eq-Asymptotic}) 
and the Laurent series ((\ref{eqLaurentKdV}), (\ref{eqLaurent-SC})).
This presents a very deep difficulty which we now describe on the following simpler example.

The Hastings-McLeod solution \cite{HML1980} of $y''=2y^3 + x y$
is a zero-parameter real solution of an irreducible second order ODE ($\PII$),
without any pole on the real line.
It admits the asymptotic behaviour
\begin{eqnarray}
& &
\left\lbrace
\begin{array}{ll}
\displaystyle{
y(x) \sim_{x \to - \infty} \left(-\frac{x}{2} \right)^{1/2},\
}\\ \displaystyle{
y(x) \sim_{x \to + \infty} \Ai(x) \sim_{x \to +\infty} 
            \frac{1}{2}\pi^{-1/2} x^{-1/4} e^{-(2/3)x^{3/2}},
}
\end{array}
\right.
\label{eqP2alpha0-Boundary-value-pb}
\end{eqnarray}
in which $\Ai$ is the Airy function,
and a local representation by the Laurent series of $\PII$
\begin{eqnarray} & &
y=\chi^{-1}\left[1 - \frac{x_0}{6} \chi^2 - \frac{1}{4} \chi^3 + u_4 \chi^4 + \frac{x_0}{72} \chi^5+O(\chi^6)  \right], \chi=x-x_0.
\end{eqnarray}
The numerical values of $x_0$
and $u_4$ which realize the matching
have never been established, at least to the author's knowledge,
and a method to perform that still has to be developed.

It would be interesting to investigate the singularity structure of the sum of the series (\ref{eqLaurent-SC})
to check how it matches the asymptotic expansion (\ref{eq-Asymptotic})
of the common solution.
Pad\'e approximants are the natural tool to do that,
they require many more terms than displayed in (\ref{eqLaurent-SC})
and there is no difficulty to compute them.

\section*{Acknowledgements}

We warmly thank Maxim Pavlov for having drawn our attention to this problem.
Relevant remarks of the referees which helped to improve the manuscript are gratefully accknowledged.


\vfill\eject

\end{document}